\documentclass[prd,aps,twocolumn,a4paper,floatfix,nofootinbib,preprintnumbers]{revtex4-1}
\usepackage[utf8]{inputenc}
\usepackage{graphicx,psfrag}
\usepackage{mathrsfs}
\usepackage{amsmath,amsfonts,amssymb}
\usepackage{multirow} 
\usepackage{diagbox}
\usepackage{comment}
\usepackage{xcolor}
\usepackage{enumerate}
\usepackage{booktabs}
\usepackage{orcidlink}
\usepackage{booktabs}
\usepackage{enumitem}
\usepackage{soul}

\usepackage{hyperref}
\hypersetup{
    colorlinks = true,
    linkcolor = {blue},
    citecolor = {blue},
    urlcolor = {blue},
    linkbordercolor = {white},
    }

\usepackage{color}
\definecolor{cyan}{rgb}{0,0.9,0.9}
\definecolor{orange}{rgb}{0.9,0.5,0}
\definecolor{magenta}{rgb}{1,0,1}
\definecolor{purple}{rgb}{0.8,0.4,0.8}
\definecolor{gray}{rgb}{0.8242,0.8242,0.8242}
\definecolor{mgreen}{rgb}{0.1,0.8,0.1}

\usepackage[normalem]{ulem}

\begin{document}

\title{Unified study of nuclear physics and dark matter constraints through gravitational-wave observations of binary neutron star mergers}

\author{Nina \surname{Kunert}$^{1,2,3}$\,\orcidlink{0000-0002-1275-530X}}
\email{nina.kunert@oca.eu}
\author{Guilherme \surname{Grams}$^{3}$\,\orcidlink{0000-0002-8635-383X}}
\author{William \surname{Newton}$^{4}$\,\orcidlink{0000-0003-3572-8299}}
\author{Edoardo \surname{Giangrandi}$^{3,5}$\,\orcidlink{0000-0001-9545-466X}}
\author{Anna \surname{Puecher}$^{3}$\,\orcidlink{0000-0003-1357-4348}}
\author{Hauke \surname{Koehn}$^{3}$\,\orcidlink{0009-0001-5350-7468}}
\author{Violetta \surname{Sagun}$^{6}$\,\orcidlink{0000-0001-5854-1617}}
\author{Tim \surname{Dietrich}$^{3,7}$\,\orcidlink{0000-0003-2374-307X}}

\affiliation{${}^1$ Université Côte d'Azur, Observatoire de la Côte d'Azur, CNRS, Laboratoire Lagrange, Nice, France}
\affiliation{${}^2$ Université Côte d'Azur, Observatoire de la Côte d'Azur, Artemis, CNRS, Nice, France}
\affiliation{${}^3$Institute for Physics and Astronomy, University of Potsdam, D-14476 Potsdam, Germany}
\affiliation{${}^4$Department of Physics and Astronomy, East Texas A\&M University, Commerce, Texas 75429-3011, USA}
\affiliation{${}^5$CFisUC, Department of Physics, University of Coimbra, Rua Larga P-3004-516, Coimbra, Portugal}
\affiliation{${}^6$Mathematical Sciences and Southampton Theory Astrophysics and Gravity (STAG) Research Centre, University of Southampton, Southampton SO17 1BJ, United Kingdom}
\affiliation{${}^7$Max Planck Institute for Gravitational Physics (Albert Einstein Institute), Am M\"uhlenberg 1, Potsdam 14476, Germany}

\date{Published in Phys.\ Rev.\ D \textbf{114}, 064056 (2026); DOI: \href{https://doi.org/10.1103/9vds-3nmb}{10.1103/9vds-3nmb}}

\begin{abstract} 
{Understanding the properties of strongly interacting matter at extreme densities is a central problem in fundamental physics, with neutron star mergers offering a natural laboratory to probe this regime. However, the complexity of the merger process complicates the interpretation of the associated gravitational-wave and electromagnetic signals. This picture becomes even more complex in the potential scenario in which dark matter accumulates around and in neutron stars, altering their structure and the associated observables. In this work, we study synthetic gravitational-wave observations of binary neutron star mergers with next-generation detectors, investigating their potential to extract both nuclear physics and dark matter constraints. We also examine how the potential presence of fermionic, non-self-interacting dark matter inside neutron stars affects the inference of nuclear empirical parameters. 
We find that combining observations can tighten constraints on nuclear empirical parameters. However, the inferred values remain sensitive to systematic modeling biases and intrinsic degeneracies among the parameters. Conversely, our analysis reveals that for non-self-interacting fermionic dark matter at mass fractions below one percent, it will be unlikely to find decisive evidence for dark matter when analyzing gravitational-wave signals. Consequently, systematic biases in nuclear empirical parameter inference potentially resulting from such a dark matter component are expected to be negligible even for observations with next-generation gravitational-wave detectors.}
\end{abstract}

\maketitle

\section{Introduction}\label{sec:intro}

\noindent
Despite the success of the standard $\Lambda$ cold dark matter model, fundamental components of the Universe, including dark matter (DM) and dark energy, remain poorly understood. Nevertheless, the majority of our knowledge is based on observations of baryonic matter (BM), although it contributes only a minor fraction to the total energy budget~\cite{Planck:2018vyg, DES:2021wwk}. In fact, for the dominant mass component, DM, its composition, interactions, and potential coupling to Standard Model particles are unknown. At the same time, the properties of strongly interacting matter across a wide range of temperatures and baryonic densities are also not well constrained \cite{Lattimer:2021emm}. 

Neutron stars (NSs) play a central role in this pursuit. In a setting where extreme gravity meets extreme density, they naturally probe the regime of strongly interacting matter. Yet the properties of the NS equation of state (EOS), which governs the thermodynamic relationship between density and pressure in this regime~\cite{Lattimer:2000nx, Lattimer:2021emm}, remain largely unconstrained~\cite{Koehn:2024set, Chatziioannou:2024tjq}. Conversely, NSs, by virtue of their long lifetimes and strong gravitational fields, constitute potential sites for DM accumulation~\cite{Bertone:2007ae, Berezinsky:2007qu, deLavallaz:2010wp, Bramante:2021dyx, Bell:2020jou, Robles:2022llu, Grippa:2024ach}. Hence, they offer, in principle, unique laboratories for studying nuclear physics and DM phenomenology in a unified setting.  

Understanding the NS EOS benefits from combining information from both terrestrial experiments and astrophysical observations, each of which probes different density regimes and offers complementary insights into the nature of dense matter \cite{Lattimer:2006xb, Hebeler:2013nza, LIGOScientific:2018cki, Pang:2022rzc, Koehn:2024set}. Experimental constraints from terrestrial facilities, such as heavy-ion collision experiments~\cite{Russotto:2016ucm}, the PREX-II~\cite{PREXII} or CREX~experiments~\cite{CREX:2022kgg}, as well as theoretical calculations of nuclear matter within the framework of chiral effective field theory ($\chi$EFT)~\cite{Lynn:2015jua,Holt:2016pjb,Drischler:2017wtt,Piarulli:2019pfq,Keller:2020qhx}, have been essential for studying the behavior of nuclear matter in regimes up to about twice the nuclear saturation density, $n_{\rm sat}\approx 0.16\,\rm{fm}^{-3}$. However, the cores of NSs extend into significantly higher density regimes, where direct experimental access is no longer possible. In these regimes, macroscopic observables such as NS masses, radii, and tidal deformabilities become key probes of the EOS, as they are sensitive to the properties of matter at supranuclear densities, typically several times the nuclear saturation density. 

Astrophysical observations such as gravitational-wave (GW) detections from binary neutron star (BNS) mergers \cite{TheLIGOScientific:2017qsa, LIGOScientific:2018cki, LIGOScientific:2018hze}, electromagnetic (EM) measurements of isolated NSs such as NICER~\cite{Riley:2019yda, Miller:2019cac, Riley:2021pdl, Miller:2021qha} and heavy pulsar measurements \cite{Demorest:2010bx, Antoniadis:2013pzd, NANOGrav:2019jur, Fonseca:2021wxt, Shamohammadi:2022ttx, NANOGrav:2023hde}, provide essential constraints in this high-density domain. 
Moreover, the first multimessenger observation of a BNS merger \cite{LIGOScientific:2017ync} through GWs (GW170817)~\cite{TheLIGOScientific:2017qsa} and EM emission (GRB~170817A and AT2017gfo)~\cite{LIGOScientific:2017zic, Goldstein:2017mmi, Coulter2017, Kasen:2017sxr, Arcavi:2017xiz, Lipunov:2017dwd, Valenti2017} allowed new constraints on the NS EOS~\cite{LIGOScientific:2018cki, Bauswein:2017vtn, Radice:2017lry, Most:2018hfd, Tews:2018iwm, Coughlin:2018miv, Dietrich:2020efo, Huth:2021bsp}. 

A systematic connection between macroscopic observational constraints and the underlying nuclear microphysics can be established through the framework of nuclear empirical parameters (NEPs). These parameters offer a compact description of the bulk properties of nuclear matter near saturation density, entering as coefficients in the Taylor expansion of the energy per particle and being derivable for a wide class of nuclear physics energy-density functionals. This makes them a useful interface between nuclear theory and experiment and has motivated recent studies aimed at constraining NEPs using astrophysical observations. Previous work has shown that the tidal deformability measurements are sensitive to the density dependence of the nuclear symmetry energy, indicating that GW observations can, in principle, probe underlying NEPs~\cite{Krastev:2018nwr, Carson:2018xri, Suleiman:2025dkn}. However, these studies also highlight substantial degeneracies among NEPs and limitations in disentangling individual parameters.

Looking ahead, next-generation GW detectors, such as the Einstein Telescope (ET) \cite{Punturo:2010zza, Freise:2008dk, Hild:2009ns, Hild:2010id, ET:2025xjr} and Cosmic Explorer (CE) \cite{LIGOScientific:2016wof, Reitze2019, Evans:2021gyd}, are expected to substantially improve sensitivity to NS EOS and thereby place more stringent constraints on NEPs. Several studies suggest that large populations of high signal-to-noise BNS detections could significantly improve the recovery of NS properties and EOS parameters, although intrinsic degeneracies among NEPs may persist \cite{Iacovelli:2023nbv, Mondal:2021vzt, Xie:2020tdo, Wouters:2025zju, Zhu:2025dea}. Additional uncertainties may emerge if the NS EOS is further altered by the possible presence of DM inside NSs. 

Astrophysical observations indicate the cosmic abundance of DM \cite{Zwicky:1933gu, Rubin:1970zza, Markevitch:2003at, Planck:2018vyg}, whereas its distribution in the galaxies, whether smooth or organized into dense subhalo structures \cite{Berezinsky:2007qu, Bramante:2021dyx}, remains an open question. However, compact objects such as NSs may accrete DM as they move through regions of enhanced DM density, where scattering interactions can facilitate capture \cite{Bell:2020jou, Berezinsky:2007qu, Bramante:2021dyx}. Over time, this process can lead to substantial DM accumulation within NSs \cite{Robles:2022llu, Bertone:2007ae, deLavallaz:2010wp}. DM, whether concentrated in the core or clustered around an NS, can modify its internal structure, affecting key properties such as mass, radius, thermal evolution, and tidal deformability \cite{Bertone:2007ae, Karkevandi:2021ygv, Ivanytskyi:2019wxd, Giangrandi:2022wht, Jockel:2023rrm, Giangrandi:2024qdb, Grippa:2024ach}. These structural modifications might impact the matter distribution and leave distinct imprints on GW signals, making BNS mergers potential observational probes for the presence and effects of DM. 

Several studies have therefore investigated whether current observations can already constrain properties of DM-admixed (DMA) NSs \cite{Mariani:2023wtv, Kumar:2024zzl,  Rutherford:2024uix, Zhang:2025rur}. These analyses suggest that, while mass-radius measurements can limit the allowed DM fraction in admixed NSs, they generally cannot unambiguously distinguish between purely baryonic and DMA configurations, highlighting the need for future precision observations.

The analysis of BNS systems offers one complementary probe of DMA NSs, as tidal interactions imprint information about their internal structure in the GW signal. For asymmetric, noninteracting fermionic DM, the magnitude of these imprints on the GW signal is governed by the DM particle mass and the accumulated DM fraction \cite{Nelson:2018xtr, Ivanytskyi:2019wxd, Konstantinou:2024ynd}. Given the current lack of constraints on both quantities, the resulting change to the tidal deformability, and hence the GW signal, is generally expected to be small. 

To further investigate such effects, recent numerical simulations have examined mergers that incorporate DM-related physics \cite{Bezares:2017mzk, Emma:2022xjs, Ruter:2023uzc, Giangrandi:2025rko}. The study of \cite{Giangrandi:2025rko} shows that the morphology of the DM can influence the merger outcome and that systems with DM halos produce GW signals that deviate from standard analytical predictions, emphasizing the need for improved waveform models which accurately capture the effect of a dilute secondary component on the tidal deformability in future parameter-estimation studies.

Detecting such subtle deviations in the tidal deformability requires the increased sensitivity of next-generation GW observatories. Using mock posteriors derived within the Fisher information matrix formalism, the study of \cite{Koehn:2024gal} showed that even with next-generation GW detectors, fermionic noninteracting DM effects on tidal deformabilities may remain difficult to disentangle due to degeneracies with EOS parameters and expected small DM fractions. These findings motivate a more comprehensive treatment that goes beyond approximate inference methods and explicitly accounts for the EOS uncertainties and DM effects in a unified framework.

In this work, we investigate the extent to which GW observations of BNS mergers with the next-generation detectors can constrain nuclear physics and DM properties. Using GW signals of simulated BNS systems analyzed with Bayesian parameter estimation, we assess the impact of different nuclear physics models on the inference of NEPs. In addition, we explore the sensitivity of these observations to fermionic, noninteracting DM by examining how its particle mass and accumulated fraction can be constrained. Finally, we systematically examine how the potential presence of DM could bias the recovery of NEPs. This approach provides the first unified assessment of the interplay between dense matter microphysics and DM effects in the inference of NEPs from BNS mergers within a realistic Bayesian inference framework.

The paper is organized as follows. In Sec.~\ref{sec:BMDMEOS_methods}, we describe the Bayesian inference framework, EOS construction, DM model, and GW data analysis setup. In Sec.~\ref{sec:results_BM}, we study nuclear physics constraints and EOS-induced biases. Section~\ref{sec:results_DM} presents results on DM parameter inference and we analyze the impact of DM admixtures on the NEP inference. Finally, we discuss our results in Sec.~\ref{Sec:Dicussion} and summarize our main findings in Sec.~\ref{sec:conclusion}.

\section{Methodology}\label{sec:BMDMEOS_methods}

\noindent
To investigate GW observations of both DMA and non-DM BNS mergers, we employ a Bayesian framework that allows us to incorporate information from nuclear and DM physics. In this section, we recap the Bayesian methodology and outline the construction of EOS sets based on different frameworks, the considered two-fluid DM framework, the construction of the DMA EOS, and the inference of the posterior distributions of nuclear and DM parameters.

\subsection{Bayesian inference}\label{Subsec:Bayesian_inference}

\noindent
Adhering to a Bayesian approach, we employ Bayes's theorem 
\begin{equation}\label{Eq:Bayestheorem}
        p(\boldsymbol{\theta} | d, \mathcal{H}) = \frac{p(d|\boldsymbol{\theta}, \mathcal{H}) p(\boldsymbol{\theta} | \mathcal{H})}{p(d|\mathcal{H})} \equiv \frac{\mathcal{L}(d|\boldsymbol{\theta}) \pi(\boldsymbol{\theta})}{\mathcal{Z}},
\end{equation}

\noindent
to obtain information on the posterior probability distributions $p(\boldsymbol{\theta} | d, \mathcal{H})$ of model parameters $\boldsymbol{\theta}$. In Eq.~(\ref{Eq:Bayestheorem}), $\mathcal{L}(d | \boldsymbol{\theta})$, $\pi(\boldsymbol{\theta})$, and $\mathcal{Z}$ refer to the likelihood, prior, and evidence, respectively. Assuming stationary Gaussian noise, the likelihood function for a GW signal $h(\boldsymbol{\theta})$ given observational data 
$d$ is expressed as~\cite{Veitch:2014wba}
\begin{equation}\label{Eq:GWLikelihood}
        \mathcal{L}_{\rm{GW}} \propto \exp\left(-\frac{1}{2}\langle d-h(\boldsymbol{\theta})|d-h(\boldsymbol{\theta})\rangle\right),
\end{equation}

\noindent
where the inner product $\langle a|b\rangle$ is defined as
\begin{equation}
        \langle a | b \rangle = 4\Re\int_{f_\textrm{low}}^{f_\textrm{high}}\frac{\tilde{a}(f)\tilde{b}^*(f)}{S_n(f)}df,
\end{equation}

\noindent
in which $\tilde{a}(f)$ represents the Fourier transform of $a(t)$, ${}^*$~denotes complex conjugation, $\Re$ denotes the real part, and $S_n(f)$ corresponds to the one-sided power spectral density of the noise. The optimal matched-filter signal-to-noise ratio (SNR) associated with a given waveform $h(t)$ is then given by
\begin{equation}
\rho_{\mathrm{opt}} = \sqrt{\langle h | h \rangle},
\end{equation}

\noindent
which quantifies the expected detection significance of a GW signal in Gaussian noise. The high-dimensional likelihood landscape is explored using nested sampling \cite{Skilling:2006}, employing the \texttt{dynesty} sampler \cite{Speagle:2019ivv} as implemented within the Bayesian inference framework \texttt{Bilby} \cite{Ashton2019}. In this study, we extend its capabilities to incorporate the inference of DM properties, operating under the assumption that BNS mergers may contain admixtures of DM.

Throughout this work, we employ the Jensen-Shannon divergence (JSD) as a symmetrized and finite measure of the dissimilarity between two probability distributions $p_1$ and $p_2$, defined as~\cite{Kullback:1951zyt, JSD_Lin}
\begin{eqnarray}\label{Eq:JSD_defintion}
    \begin{aligned}
    D_{\rm JS}(p_1(x) || p_2(x)) = \frac{1}{2} \Biggl[ &\sum_x p_1(x) \ln \Bigl( \frac{p_1(x)}{m(x)} \Bigr) +\\ 
    &\sum_x p_2(x) \ln \Bigl( \frac{p_2(x)}{m(x)}\Bigr) \Biggr],
    \end{aligned}
\end{eqnarray}
in which $m(x) = 0.5(p_1(x) + p_2(x))$ denotes the mixture distribution. The JSD is symmetric and bounded, $0 \leq D_{\rm JS} \leq \ln 2$, corresponding to identical ($p_1 = p_2$) and nonoverlapping ($p_1p_2 = 0$) distributions, respectively. For convenience, we report the normalized divergence $D_{\rm JS}/\ln 2 \in [0, 1]$, expressed in bits. 

\subsection{Nuclear empirical parameters}\label{sec:NEPs_description}

\noindent
NEPs provide a convenient phenomenological description of the bulk properties of nuclear matter near its equilibrium state, as they encode the density dependence of the energy per particle by expanding the energy of nuclear matter around saturation density $n_{\rm sat}$. In this framework, the energy per particle of symmetric nuclear matter $e_{\rm SM}$ and the symmetry energy $e_{\rm sym}$ are written as Taylor expansions in the dimensionless parameter 
\begin{equation}
x := \frac{n - n_{\rm sat}}{3 n_{\rm sat}} \, ,
\end{equation}
\noindent
leading to 
\begin{eqnarray}
\label{eq:sat}
&&\hspace{-1cm}e_{\rm SM}(n) = E_{\rm sat} + K_{\rm sat} \frac{x^2}{2} + Q_{\rm sat} \frac{x^3}{3!} \nonumber \\
&&\hspace{2cm} + Z_{\rm sat} \frac{x^4}{4!} + \dots \, , \\
\label{eq:sym}
&&\hspace{-1cm}e_{\rm sym}(n) = E_{\rm sym} + L_{\rm sym} x + K_{\rm sym} \frac{x^2}{2}+ Q_{\rm sym} \frac{x^3}{3!} \nonumber \\ 
&&\hspace{2cm} + Z_{\rm sym} \frac{x^4}{4!} + \dots \, .
\end{eqnarray}

The coefficients appearing in Eqs.~(\ref{eq:sat}) and (\ref{eq:sym}) are defined with respect to derivatives of the energy per particle with respect to the baryon number density and are evaluated at $n = n_{\rm sat}$. In the isoscalar channel, $E_{\rm sat}$ denotes the energy per particle of symmetric nuclear matter at saturation density. The higher-order density dependence is governed by the incompressibility $K_{\rm sat}$, the skewness parameter $Q_{\rm sat}$, and the kurtosis parameter $Z_{\rm sat}$, which probe increasingly higher-order derivatives of the energy per particle with respect to density. In the isovector channel, the symmetry energy at saturation is given by $E_{\rm sym}$. Its density dependence is quantified by the slope parameter $L_{\rm sym}$, the curvature $K_{\rm sym}$, the skewness $Q_{\rm sym}$, and the kurtosis $Z_{\rm sym}$, which control the behavior of neutron-rich matter at and above saturation density. 

Together, these coefficients define the NEPs and characterize the bulk properties of nuclear matter that govern the pressure-density relation of dense matter. As a result, variations in NEPs play a central role in determining the NS EOS and the associated macroscopic observables, such as masses, radii, and tidal deformabilities.

Within the metamodel framework \cite{Margueron:2017eqc, Margueron:2017lup}, NEPs enter explicitly as model parameters, providing a flexible and transparent way to explore nuclear physics uncertainties. In other EOS approaches, NEPs can be derived \emph{a posteriori} from the underlying model parameters, enabling a unified comparison across different nuclear physics frameworks. In this way, NEPs provide a direct link between terrestrial laboratory measurements and theoretical descriptions of matter under the extreme conditions encountered in NSs.

\subsection{Employed neutron star equations of state}\label{Subsec:BMEOSsets}

\noindent
Throughout this work, we employ three distinct EOS models to describe strongly interacting matter, each capturing different aspects of nuclear microphysics and high-density behavior. Below, we outline the main ideas for the three considered models, while we provide more details on the EOS construction in the Appendix.

\begin{enumerate}
    \item \textbf{Metamodel (MM) approach with speed-of-sound extension (MM-SS):} This approach \cite{Margueron:2017eqc, Tews:2018iwm, Somasundaram:2021clp, Margueron:2017lup, Komoltsev:2023zor, Koehn:2024set} parametrizes the nuclear EOS in terms of NEPs through a Taylor expansion of the energy per particle around saturation density. At higher densities, additional flexibility is introduced via a speed-of-sound (SS) extension, which allows for extra degrees of freedom. This model enables us to study a wide range of high-density behaviors while remaining agnostic about the microscopic composition and is characterized by eight NEPs, one SS breakpoint parameter, and additional fixed parameters for the SS parametrization. 
    
    \item \textbf{Metamodel approach with unified crust-core EOS (MM):} The second EOS family is based on the same metamodel framework but assumes purely nucleonic matter at all densities and employs a unified treatment of the crust and core. In this construction, NEPs explicitly parametrize the EOS throughout the entire star, from the crust to the inner core, providing a consistent description of nuclear matter across all density regimes. The unified MM employs ten NEPs to describe purely nucleonic matter across the crust and core, with all crust model parameters fixed~\cite{Grams:2022a}.
    
    \item \textbf{Skyrme model:} The third EOS family is based on nonrelativistic density functional theory using extended energy-density functionals (EDFs) \cite{Newton:2011dw, Davesne:2015lca,zhangExtendedSkyrmeInteractions2016, Newton:2021rni,Balliet:2020nsh}. The Skyrme model employed here can be written as linear combinations of the underlying coupling constants for fixed density-dependent exponents and allows independent variation of the NEPs up to and including the $Q_{\rm sat}$ and $Q_{\rm sym}$ orders, one fewer than in the MM expansion. As an EDF framework, it supports a unified crust-core description and enables direct connections to laboratory observables such as neutron skin thicknesses of heavy nuclei~\cite{PREXII,CREX:2022kgg}. 
\end{enumerate}
\noindent
Throughout this work, we use the term ``non-DM EOS models'' to collectively refer to all three baseline NS matter EOS considered, including both purely baryonic EOS and EOS that allow for nonbaryonic degrees of freedom, e.g., through phase transitions at high densities. When referring specifically to EOS that assume purely BM at all densities, we explicitly use the term baryonic EOS models.

\begin{figure}[h!]
    \centering
    \includegraphics[width=\columnwidth]{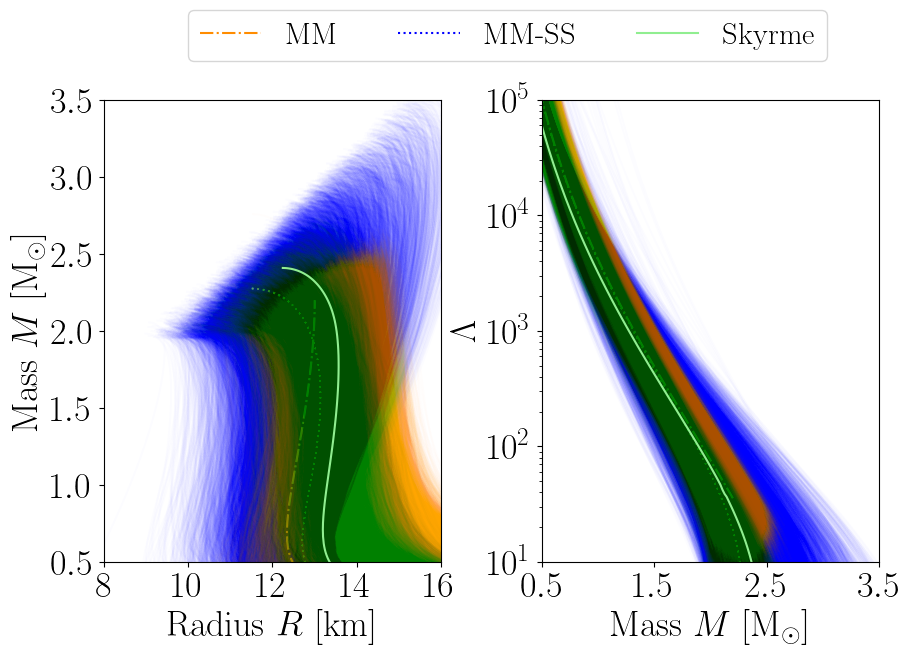}
    \caption{Overview of the employed EOS sets illustrating the macroscopic NS properties. The left panel depicts the mass-radius relationship, while the right panel shows the tidal deformability-mass plane. The MM-SS EOS set is represented in blue, the MM EOS set in orange, and the Skyrme EOS set is shown in green. Injected EOS within each set are shown using the same color scheme.}
    \label{fig:MRL_EOSsets}
\end{figure}

\begin{figure*}
    \centering
    \includegraphics[width=\textwidth]{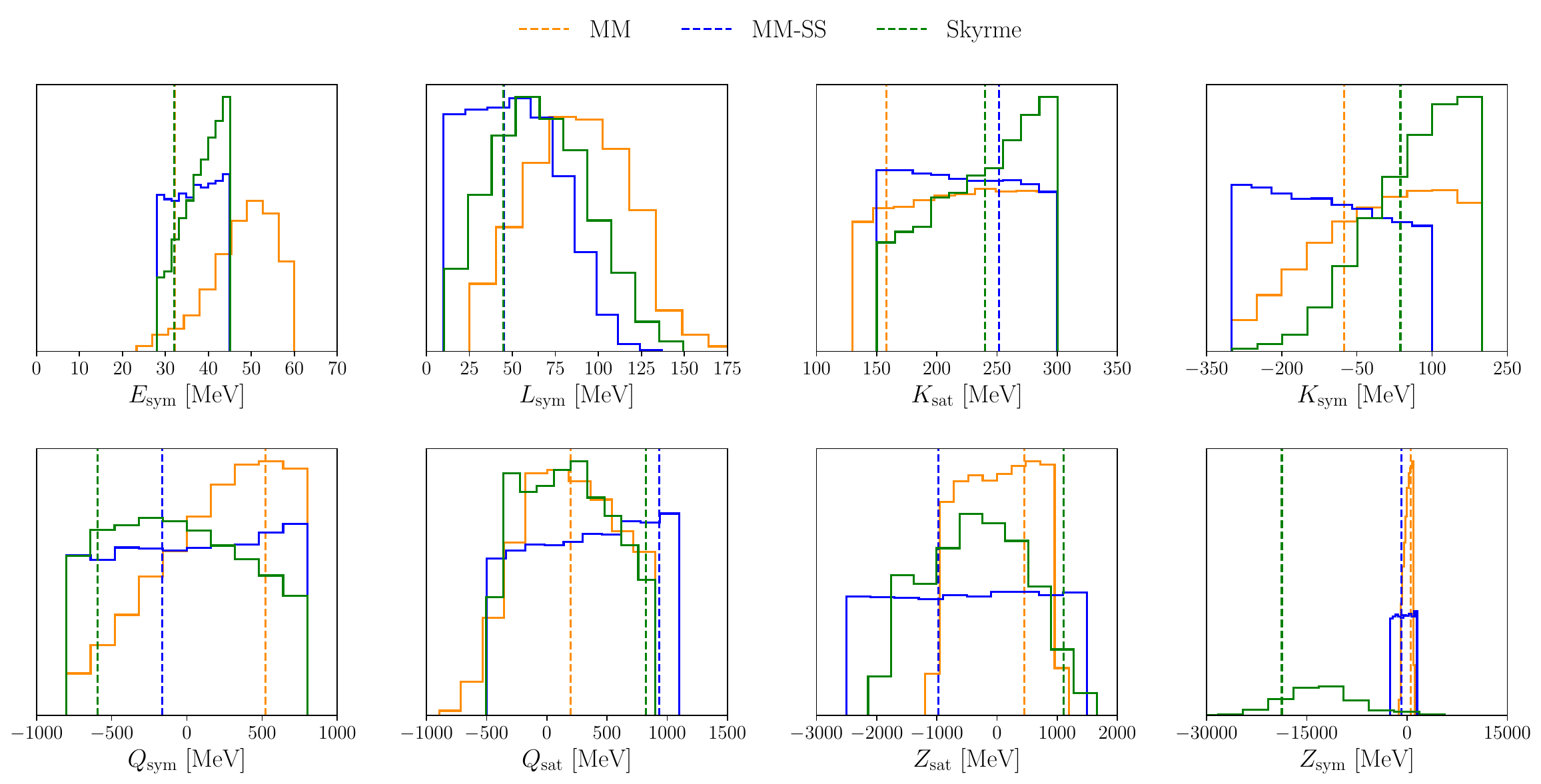}
    \caption{Normalized prior distributions of NEPs for the different EOS described in Sec.~\ref{Subsec:BMEOSsets}. The selected EOS injections are shown as vertical dashed lines for each of the models and are motivated through findings in~\cite{Koehn:2024set}.}
    \label{fig:injEOS_basedon_Nucparams}
\end{figure*}

Figure~\ref{fig:MRL_EOSsets} illustrates the distinct macroscopic predictions of the EOS families through their mass-radius ($M$-$R$) and mass-tidal deformability ($M$-$\Lambda$) relations.
The corresponding prior distributions of the NEPs are shown in Fig.~\ref{fig:injEOS_basedon_Nucparams}. The $M$-$\Lambda$ plane shows systematic differences among the EOS families. The MM and Skyrme models populate narrower, overlapping regions, with the MM EOS allowing moderately larger tidal deformabilities at intermediate masses. The Skyrme EOS are confined to comparatively smaller $\Lambda$ values, consistent with their more restricted stiffness arising from the lower upper bounds of the $E_{\rm sym}$ and $L_{\rm sym}$ priors. The MM-SS EOS span a broader range in $\Lambda$ and extend to lower values at fixed mass, reflecting the combined effect of wider NEP priors and the additional freedom introduced by the SS extension. 

One notable feature of the priors shown in Fig.~\ref{fig:injEOS_basedon_Nucparams} is that the two unified crust-core models (MM and Skyrme) show a prior that favors higher values of $E_{\rm sym}$ and suppresses lower values.  In nonunified models the crust is
fixed, so the NEP prior used for the NS core is unaffected. In a unified model, by contrast, the crust contains the same NEPs as the core and the compressible liquid drop model (CLDM) must admit a solution: clusters in mechanical, chemical, $\beta$, and virial equilibrium with a dripped neutron fluid~\cite{Newton:2011dw,Grams:2022a}. This requires pure neutron matter to remain unbound at subsaturation densities, with positive and increasing pressure. Combinations of low $E_{\rm sym}$ and high $L_{\rm sym}$ instead produce neutron matter instabilities at crust densities, and the CLDM equilibrium equations then admit no solution. Such models are known to be filtered out of viable crust calculations~\cite{Newton:2021rni}. There are far more EOS with this feature at low values of $E_{\rm sym}$ than high values.

\subsection{Dark matter model}\label{Sec:DM_model}
\noindent
In this study, DM is modeled as a relativistic Fermi gas, assuming minimal coupling with the Standard Model particles. At zero temperature, the pressure $p_{\rm DM}$ of DM is given as \cite{Narain:2006kx, Nelson:2018xtr,  Ivanytskyi:2019wxd}

\begin{equation}
\label{Eq:DMEOS}
\begin{aligned}
    p_{\rm DM} &= \frac{g_{\rm DM}}{48\pi^2}[\mu_{\rm DM} k_{\rm DM} (2\mu_{\rm DM}^2-5m_{\chi}^2)\\
    &+ 3m_{\chi}^4\ln(\frac{\mu_{\rm DM}+k_{\rm DM}}{m_{\chi}})],
\end{aligned}
\end{equation}
where $\mu_{\rm DM}$, $g_{\rm DM}=2$, $n_{\rm DM}=\partial p_{\rm DM} / \partial \mu_{\rm DM}$, and $k_{\rm DM}=\sqrt{\mu_{\rm DM}^2-m_{\chi}^2}\theta(\mu_{\rm DM}-m_{\chi})$ are the DM chemical potential, degeneracy factor, number density, and Fermi momentum, respectively. The energy density of the DM component is found from the thermodynamic identity $\varepsilon_{\rm DM}=\mu_{\rm DM} n_{\rm DM} -p_{\rm DM}$~\cite{Nelson:2018xtr}.

To model the DMA NSs, we adopt the two-fluid approach, widely used in previous studies, e.g.,~\cite{Kodama_1972, Henriques:1989ar, Sandin:2008db, Ivanytskyi:2019wxd,Sagun:2023rzp, Konstantinou:2024ynd}, in which the two-fluid Tolman-Oppenheimer-Volkoff (TOV) equations for a DMA NS in hydrostatic equilibrium can be formulated as~\cite{Ivanytskyi:2019wxd}
\begin{eqnarray}
\label{Eq:TwoFluidTOV}
    \frac{dp_j}{dr} 
    &=&- \frac{(\epsilon_j + p_j) (M + 4\pi r^3 p)}{r^2 (1-2M/r)},\\
    \frac{dM_j}{dr} 
    &=& 4\pi r^2 \epsilon_j(r),
    \label{Eq:TwoFluidMass}
\end{eqnarray}
where the subscript index $j$ in Eqs.~(\ref{Eq:TwoFluidTOV}) and~(\ref{Eq:TwoFluidMass}) represents the different fluid components, namely DM and BM. Here, the $\epsilon_j$ and $p_j$ are the energy density and pressure of each of the components. Accordingly, the total gravitational mass and pressure are given by $M(r) = M_{\rm BM}(r) + M_{\rm DM}(r)$ and $p(r) = p_{\rm BM}(r) + p_{\rm DM}(r)$, respectively, where $r$ denotes the radial coordinate from the center of the NS. The two components are thus coupled only through the total mass and pressure entering the right-hand side of Eq.~(\ref{Eq:TwoFluidTOV}). A derivation of this system from the Einstein field equations is given in the Appendix of Ref.~\cite{Ivanytskyi:2019wxd}. For fixed values of central pressures, we integrate Eqs.~(\ref{Eq:TwoFluidTOV}) and (\ref{Eq:TwoFluidMass}) outward from the center of the star. Each component has its own radius, $R_j$, at which its pressure vanishes, $p_j (R_j) =0$. Beyond that radius the component no longer contributes, while the other one continues to be integrated until its own surface is reached.

The DM model is characterized by two parameters: the DM particle mass $m_{\chi}$ and the DM mass fraction $f_{\chi}$. The DM mass fraction,
\begin{equation} 
    f_{\chi} = \frac{M_{\rm DM} (R_{\rm DM})}{M}, 
\end{equation}
quantifies the proportion of DM mass relative to the total NS mass enclosed within its radius and is hereafter referred to as DM fraction. The choice of these DM parameters yields two distinct configurations: the core configuration, where DM is entirely confined within the NS, and the halo configuration, where DM extends beyond the BM component, forming a diffuse halo surrounding the NS. Consequently, these configurations will affect the NS properties such as mass, radius, and tidal deformability, as shown in~\cite{Ivanytskyi:2019wxd, Giangrandi:2022wht}. 

While core configurations generally lead to a reduction of the maximum mass, halo configurations result in an increase in the maximum NS mass at higher DM fractions. As shown in Ref.~\cite{Giangrandi:2025rko}, extended DM halos may overlap during the late stages of BNS inspirals, and DM-induced modifications of tidal deformabilities would need to be taken into account to accurately model GW signals from such systems. However, existing gravitational waveform models do not include these effects. For this reason, we focus on core configurations and assume that both NSs in a binary have the same DM fraction.  

\subsection{Construction of dark-matter-admixed neutron stars}\label{subsec:DMEOS_construction}

\noindent
In Sec.~\ref{sec:results_DM}, we investigate BNS mergers incorporating DM admixtures with the objectives of (i) assessing the ability of GW observations to constrain the underlying DM properties and (ii) evaluating whether the presence of an additional DM component introduces systematic biases in the inference of baryonic NEPs. In this context, DMA BNS mergers refer to NSs composed of both baryonic and DM components.

Building on the microphysical descriptions of pure BM presented in Sec.~\ref{Subsec:BMEOSsets}, we extend the baryonic analysis by incorporating the noninteracting Fermi gas model described in Sec.~\ref{Sec:DM_model}, which allows us to include DM effects while preserving the key physical characteristics of the system. To obtain NS masses, radii, and tidal deformabilities that account for DM effects, we construct DMA EOS based on the MM and Skyrme models introduced in Sec.~\ref{Subsec:BMEOSsets}. This choice is both physically motivated and supported by computational considerations. The computational cost of our analysis is driven by the large number of Bayesian samples, the high dimensionality of the parameter space, and the repeated TOV integrations required for DM-admixed NS configurations.\footnote{Concerning the MM-SS, the inclusion of DM effects is not considered, since this parametrization is designed to emulate hadron-quark phase transitions and other nonbaryonic degrees of freedom, which would introduce additional sources of degeneracy and hinder an unbiased identification of DM signatures.}

For both the MM and Skyrme EOS frameworks, a representative ensemble of 5000 baryonic EOS is randomly selected. For each EOS within this ensemble, the two-fluid TOV and tidal Love equations are solved over a predefined grid of DM particle masses and mass fractions. This procedure follows the approach in~\cite{Koehn:2024gal}, with refinements introduced in the present analysis. 

The grid of DM mass fractions is defined as
\begin{equation}
\begin{aligned}
f_\chi \,\, \rm{in} \,\, \% = 
\{&0.01,\, 0.015,\, 0.023,\, 0.035,\, 0.053,\, 0.081,\, 0.123,\, \\
  &0.187,\, 0.285,\, 0.433,\, 0.658,\, 1.0\},
\end{aligned}
\end{equation}
in which the values are log-uniformly spaced to sample the range $0.01\%$--$1\%$. The lower bound, $f_{\chi} = 0.01\%$, is motivated by estimates from Ref.~\cite{Ivanytskyi:2019wxd}, which suggest that NSs in the densest galactic regions may accumulate up to this fraction of their mass in the form of DM, while smaller admixtures are expected to have a negligible impact on stellar structure. Higher DM mass fractions could, in principle, arise under exceptional conditions, such as the rapid accretion of DM during a NS’s passage through dense primordial subhalo regions or compact DM clumps~\cite{Erickcek:2011us, Buckley:2017ttd, Bramante:2021dyx}. The upper bound, $f_{\chi} = 1\%$, is adopted as an upper benchmark to bracket potentially observable effects. For significantly larger DM fractions, Ref.~\cite{Ellis:2018bkr} demonstrated that a DM mass fraction of about $5\%$ leads to a pronounced reduction in the radius-mass relation, excessively softening the EOS and conflicting with observations of $\sim2M_{\odot}$~pulsars and GW constraints from GW170817. The chosen interval thus encompasses the astrophysically plausible regime in which the DM effects are appreciable yet remain consistent with current observational bounds.

Concerning the choice of DM particle masses, we define a log-uniformly spaced grid 
\begin{equation}
\begin{aligned}
m_\chi \,\, \rm{in} \,\, \rm{MeV} = 
\{&300,\, 370,\, 456,\, 562,\, 693,\, 854,\, 1053,\, \\
  &1299,\,1601,\, 1974,\, 2433,\, 3000\},
\end{aligned}
\end{equation}
\noindent
to sample the range $300$--$3000$~MeV. 

Previous studies indicated that light fermionic DM with $m_{\chi} \lesssim 170$~MeV forms an extended halo around the NS~\cite{Ivanytskyi:2019wxd}, which overlaps during the late inspiral phase. Such configurations are not accounted for in existing gravitational-waveform models~\cite{Ruter:2023uzc}. In contrast to Ref.~\cite{Koehn:2024gal}, we adopt a lower DM particle mass bound of $300$~MeV, as for $m_{\chi} \sim 170$~MeV we still find the formation of halo configurations, which are excluded from the present analysis. An upper limit of $m_{\chi} = 3000$~MeV is adopted as a conservative bound, since for heavier DM particles, even subpercent DM mass fractions produce excessively compact cores that significantly soften the EOS, lowering the maximum mass and tidal deformability below the limits set by observations. 

Solving the TOV and tidal Love equations for the MM and Skyrme subsets yields a corresponding family of DMA configurations characterized by DM-modified NS masses, radii, and tidal deformabilities. These solutions are precomputed for discrete values of the EOS, DM particle mass, and DM fraction and are subsequently used to assign DM-modified tidal deformabilities, which enter the gravitational waveform modeling. 

The stability of the DMA configurations is ensured as follows. For a two-fluid star the equilibrium configurations depend on two central pressures, $p_{\rm BM}(0)$ and $p_{\rm DM}(0)$. In our construction the central DM chemical potential is fixed at each step by requiring that the target DM fraction be reproduced, so that the configurations at given $m_{\chi}$ and $f_{\chi}$ form a single sequence parametrized by the central baryonic chemical potential alone. Along this sequence we retain only configurations satisfying $\partial M/\partial p_c > 0$, i.e., we truncate at the first turning point of the total gravitational mass $M = M_{\rm BM} + M_{\rm DM}$. Configurations violating causality are likewise discarded. For DMA NSs at fixed DM fraction, radial oscillation analyses \cite{Routaray:2022utr} have found the frequency of the fundamental mode to vanish at this turning point. 
For two-fluid stars, the stability boundary can more generally be characterized by the critical curve of the particle numbers in the two-dimensional space spanned by the two central densities \cite{Caballero:2024qtv}, which does not necessarily coincide with the mass turning point. This does not affect our results, since all configurations entering the analysis lie below the maximum-mass configuration of the respective EOS, while the employed EOS sets are filtered to support maximum masses of at least $\sim1.99 M_{\odot}$ (Skyrme) and to remain causal up to $1.8 M_{\odot}$ (MM).

\subsection{Employed gravitational-wave data and detector setup}\label{subsec:Employed_data}

\noindent
This study explores GW signals from both baryonic and DMA BNS mergers, with a focus on next-generation GW observatories such as the ET and CE. To generate synthetic GW data, we adopt a triangular ET configuration with a $10$~km arm length, positioned in Limburg \cite{Branchesi:2023mws, Maggiore:2019uih, ET_D_PSD}, and set a minimum frequency of $f_{\rm min} = 5$~Hz. For the analysis of DM constraints, we additionally include CE \cite{LIGOScientific:2016wof, Reitze2019, Evans:2021gyd, CE_PSD} with an arm length of $40$~km located in Hanford, forming a joint ET+CE detector network and adopting the same minimum frequency cutoff. Four different BNS configurations are simulated and analyzed (see Table~\ref{tab:BNS_events_params}).

\begin{table}[t]
\centering
\begin{tabular}{lcccc}
\hline
\hline\\
\textbf{Parameter} & \textbf{Event A} & \textbf{Event B} & \textbf{Event C} & \textbf{Event D} \\[0.3\normalbaselineskip]
\hline
\addlinespace
$\mathcal{M}_c [M_{\odot}]$ & 1.19 & 1.32 & 1.28 & 1.41 \\[0.15\normalbaselineskip]
$q$                         & 0.96 & 0.97 & 0.97 & 0.67 \\[0.15\normalbaselineskip]
$\chi_{z, 1}$                       & 0.00 & 0.03 & 0.02 & 0.04 \\[0.15\normalbaselineskip]
$\chi_{z, 2}$                       & 0.00 & 0.05 & 0.04 & 0.04 \\[0.15\normalbaselineskip]
$D_L$~[Mpc]                       & 40.00 & 609.45 & 41.20 & 116.09 \\[0.15\normalbaselineskip]
$\delta$~[rad]                    & -0.41 & 2.72 & 2.14 & 2.02 \\[0.15\normalbaselineskip]
$\alpha$~[rad]                    & 3.45 & 5.20 & 1.18 & 6.28 \\[0.15\normalbaselineskip]
$\cos \theta_N$       & -0.90 & 0.95 & 0.84 & 0.95 \\[0.15\normalbaselineskip]
$\psi$~[rad]             & 0.87 & 0.69 & 3.75 & 4.69 \\[0.15\normalbaselineskip]
$\phi_c$~[rad]             & 1.22 & 0.29 & 2.04 & 0.88 \\[0.15\normalbaselineskip]
$\rho^{\rm ET}_{\rm opt}$       & $\sim 643$ & $\sim123$ & $\sim813$ & $\sim573$ \\
\addlinespace
\hline
\hline
\end{tabular}
\caption{Overview of the BNS merger configurations simulated in this study along with their key parameters. The quantity $\mathcal{M}_c$ denotes the source-frame chirp mass, while tidal deformabilities are determined by the respective injected EOS, as outlined in Sec.~\ref{subsec:Employed_data}, and are therefore omitted for brevity. The parameter $\rho^{\rm ET}_{\rm opt}$ refers to the optimal SNR as would be detected with a triangular ET design.}\label{tab:BNS_events_params}
\end{table}

\noindent
Event A is constructed using system parameters inferred from the multimessenger Bayesian analysis of the observed GW signal GW170817 and its EM counterparts GRB170817A and AT2017gfo presented in \cite{Pang:2022rzc}. The other BNS events B–D are drawn from a BNS catalog generated in~\cite{Koehn:2024set}. 

For each EOS model, the choice of injected EOS is guided by independent constraints on the symmetry energy and its slope reported in~\cite{Koehn:2024set}, which favor values of $E_{\rm sym} \approx 32$ and $L_{\rm sym} \approx 50$~MeV. Importantly, this selection does not imply that the remaining NEPs take identical values across different EOS sets. Rather, each EOS framework leads to differences in the higher-order density dependence and high-density behavior of the EOS. For each EOS set, tidal deformabilities are injected consistently with the corresponding selected EOS for all four BNS events.

All simulations of GW observations are conducted under the assumption of aligned-spin BNS configurations, utilizing the waveform model \texttt{IMRPhenomD\_NRTidalv2}~\cite{Husa:2015iqa, Khan2016, Taracchini:2013rva, Dietrich:2019kaq}. For all simulated BNS events, we adopt broad and minimally informative priors on the intrinsic and extrinsic source parameters. The component masses are restricted to $0.5\text{–}3M_\odot$, with a uniform prior on the mass ratio $q \in [0.125,1]$, while the detector-frame chirp mass is assigned a narrow uniform prior centered on the injected value with a width of $0.1M_\odot$. Standard isotropic priors are used for sky location, orientation, polarization, and phase, and the prior on tidal deformabilities is determined by the respective EOS set. Since BNS event A is constructed from a previous multimessenger analysis, whereas events B–D are drawn from a synthetic catalog, the prior choices differ slightly between these cases. In particular, aligned-spin magnitudes are restricted to $\chi_{z_{1,2}} \leq 0.15$ for event~A and to $\chi_{z_{1,2}} \leq 0.05$ for events~B–D. Similarly, the luminosity distance prior follows a uniform-in-comoving-volume distribution with an upper bound of $75~\mathrm{Mpc}$ for event A and $1000~\mathrm{Mpc}$ for events B–D. To reduce computational costs during the sampling, we employ the parameter-estimation acceleration technique proposed in~\cite{Morisaki:2021ngj}, utilizing a multibanding accuracy factor of $L=50$.

\subsection{Inference of nuclear and dark matter properties}\label{Subsec:Inferring_NucDM_params}

\noindent
To infer the NEPs, we jointly sample binary parameters and EOS, thereby incorporating nuclear physics-informed tidal information directly into GW parameter estimation. For a particular EOS set, each EOS is assigned a discrete EOS identifier (hereafter referred to as ``EOS ID''), which uniquely specifies the underlying nuclear microphysics parameters and, in particular, the associated set of NEPs. 

Under the assumption that NSs in the binary can be described with purely baryonic EOS or EOS that allow for nonbaryonic degrees of freedom at high densities, the BNS parameters are defined as
\begin{equation}
    \begin{aligned}
    \boldsymbol{\theta}_{\rm BNS} = \{&  M_i, \chi_{z, i}, \Lambda_i(M_i; {\rm EOS}),\\ & D_L, t_c, \phi_c, \alpha, \delta, \iota, \psi \},
    \end{aligned}
\end{equation}
where $i = 1,2$ indexes the two NSs of the binary. $M_i$ denotes the individual masses, $\chi_{z, i}$ the aligned-spin component, $D_L$ the luminosity distance, $t_c$ the coalescence time, $\phi_c$ the phase, $(\alpha, \delta)$ the sky location, $\iota$ the inclination angle, and $\psi$ is the polarization angle. 
Importantly, as mentioned, the tidal deformabilities are not sampled as independent parameters, but are instead obtained, for each sample of component masses and EOS ID, by interpolation from precomputed mass–tidal deformability relations. Since each EOS ID maps uniquely to a specific set of NEPs, this approach allows posterior distributions of the NEPs to be reconstructed during postprocessing from the EOS posterior samples.

In the presence of DM, NSs are modeled within a two-fluid framework in which both BM and DM contribute to the stellar structure. In this analysis, only purely baryonic EOS approaches are employed, and EOS with SS extensions are excluded. The total gravitational mass is therefore given by \(M = M_{\rm BM} + M_{\rm DM}\), and the sampling space is extended to include the DM particle mass and mass fraction \((m_\chi, f_\chi)\). The corresponding parameter vector for DMA BNS mergers becomes
\begin{equation}\begin{aligned}\label{Eq:DMBNS_params}
    \boldsymbol{\theta}^{\rm DMA}_{\rm BNS} = \{ & M_i, \chi_{z, i}, \Lambda^{\rm DMA}_i(M_{i}; f_{\chi}, m_{\chi}, {\rm EOS}), \\ & D_L, t_c, \phi_c, \alpha, \delta, \iota, \psi \},
    \end{aligned}
\end{equation}
where EOS refers to the baryonic EOS ID used to construct the corresponding DMA configurations for different choices of DM fractions and particle masses (cf.~Sec.~\ref{subsec:DMEOS_construction}). 

For each sample of the NS masses, baryonic EOS, and DM parameters \((m_\chi, f_\chi)\), the corresponding DMA tidal deformabilities are obtained by interpolating precomputed solutions of the two-fluid TOV and tidal perturbation equations. Since gravitational waveform models that explicitly incorporate DM effects are currently unavailable, DM admixtures are accounted for by replacing the baryonic tidal deformabilities with their two-fluid counterparts in otherwise standard BNS waveform models.

\section{Studying nuclear physics constraints and biases}\label{sec:results_BM}

\subsection{Quantifying the constraints for each equation-of-state model}\label{subsec:results_BM_NucConstraints}

\begin{figure*}[t]
    \centering
    \includegraphics[width=\textwidth]{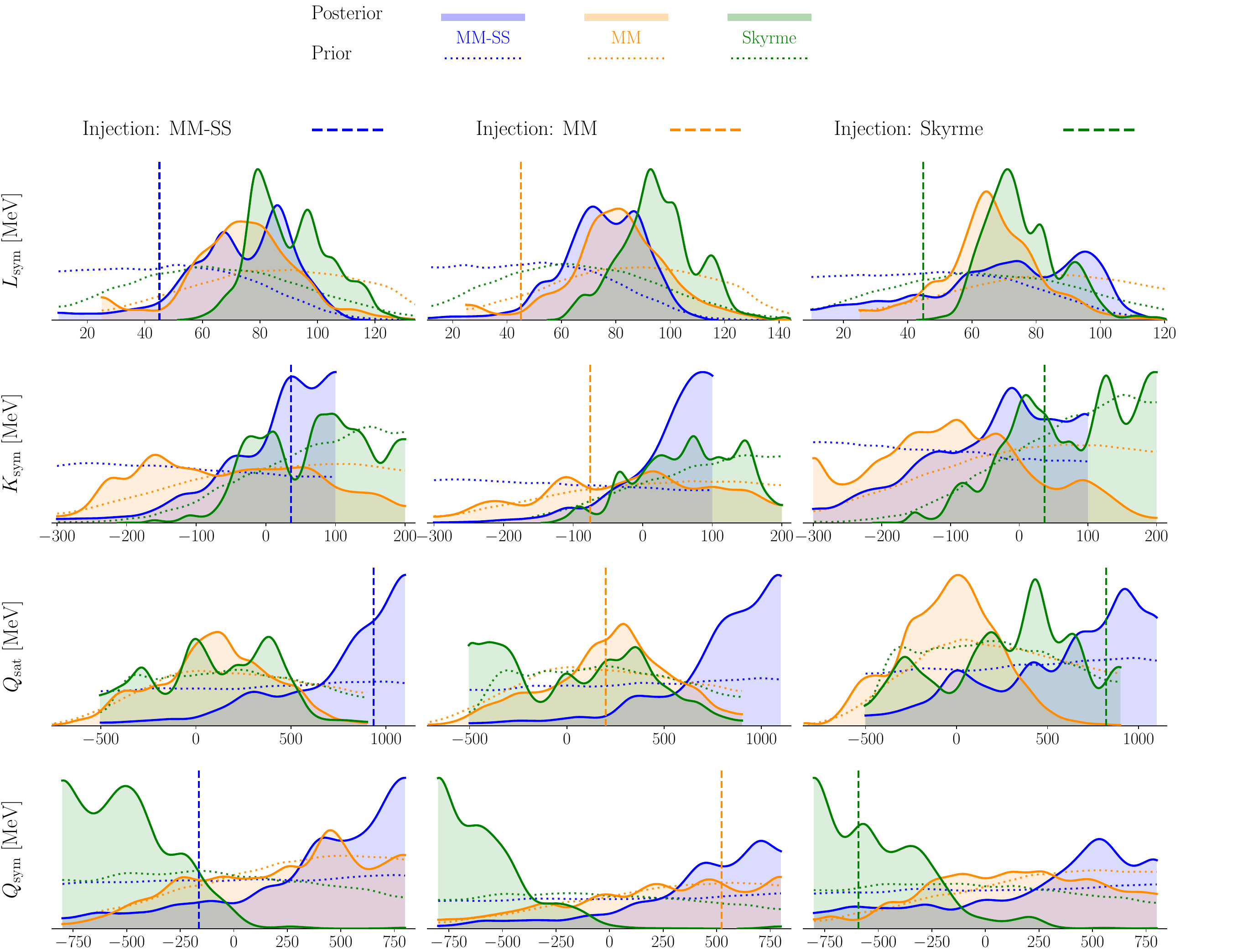}
    \caption{Posterior distributions of selected NEPs obtained when injecting with MM-SS (left column), MM (middle column), and Skyrme (right column), while recovering each injection with MM-SS (blue), MM (orange), and Skyrme (green) EOS sets, respectively. The posterior distributions represent the combined posterior across the four simulated BNS events~A–D as shaded regions and the prior is shown for each EOS as dotted lines. The injections of NEPs are color-coded, respectively, and are shown as dashed lines.}
    \label{fig:Sysbiases_complete}
\end{figure*}

\noindent
In this section, we assess the extent to which NEPs can be constrained when the injection and recovery are performed within the same EOS framework. We consider four BNS events A–D simulated with a triangular ET configuration, considering both single-event and combined-event posteriors. The associated information gain is quantified via the JSD between the prior and posterior distributions. Our results are reported in Table~\ref{JSDs_all_NUCmodels} and in Fig.~\ref{fig:Sysbiases_complete}. 

For individual events, GW observations generally do not yield informative constraints on the NEPs across all EOS models considered. An exception is found for the Skyrme framework, where the symmetry energy slope shows a modest information gain for the highest SNR event. Motivated by this limited single-event sensitivity, we examine how the constraints evolve when information from all four events is combined. The combined analysis reveals that the degree of information gain is strongly model dependent.

\begin{table}[t]
\centering
\renewcommand{\arraystretch}{1.4}
\begin{footnotesize}
\resizebox{1.0\linewidth}{!}{%
\begin{tabular}{l|cc|cc|cc}
\hline
\multicolumn{7}{c}{\textbf{JSD $D_{\rm JS}$ (in bits)}} \\
\hline
\textbf{Nuclear} 
& \multicolumn{2}{c|}{\textbf{MM-SS}} 
& \multicolumn{2}{c|}{\textbf{MM}} 
& \multicolumn{2}{c}{\textbf{Skyrme}} \\
\textbf{parameter} 
& Ev.~C & Com. 
& Ev.~C & Com. 
& Ev.~C & Com. \\
\hline
$L{_\mathrm{sym}}$ [MeV] & 0.03 & 0.16 & 0.02 & 0.10 & 0.16 & 0.27 \\
$E{_\mathrm{sym}}$ [MeV] & 0.01 & 0.01 & 0.01 & 0.01 & 0.05 & 0.03 \\
$K{_\mathrm{sat}}$ [MeV] & 0.02 & 0.09 & 0.01 & 0.03 & 0.04 & 0.02 \\
$K_{\mathrm{sym}}$ [MeV] & 0.03 & 0.21 & 0.01 & 0.02 & 0.03 & 0.02 \\
$Q{_\mathrm{sat}}$ [MeV] & 0.03 & 0.16 & 0.01 & 0.03 & 0.04 & 0.03 \\
$Q{_\mathrm{sym}}$ [MeV] & 0.02 & 0.09 & 0.01 & 0.01 & 0.07 & 0.23 \\
$Z{_\mathrm{sym}}$ [MeV] & 0.02 & 0.01 & 0.01 & 0.02 & 0.04 & 0.02 \\
$Z{_\mathrm{sat}}$ [MeV] & 0.01 & 0.02 & 0.01 & 0.02 & 0.04 & 0.05 \\
\hline
\end{tabular}
}
\end{footnotesize}
\caption{JSD values for the three models considered: MM-SS, MM, and Skyrme, evaluated using Eq.~(\ref{Eq:JSD_defintion}) based on their respective prior and posterior distributions. The left columns report the JSD inferred from the highest SNR event, C, whereas the right columns correspond to the JSD when comparing the prior to the combined posterior of all four BNS events A--D (cf.~Fig.~\ref{fig:Sysbiases_complete}).}
\label{JSDs_all_NUCmodels}
\end{table}

\noindent
\textit{Inference with the MM-SS model:} Within the MM-SS framework, the combined BNS dataset leads to a moderate information gain for parameters that govern the density dependence of the symmetry energy. In particular, the symmetry energy slope and curvature show moderate information gain in the combined dataset with JSD values of $0.16$ and $0.21$, while the skewness of symmetric nuclear matter also displays a non-negligible sensitivity. In contrast, the symmetry energy $E_{\mathrm{sym}}$ and higher-order derivatives $Q_{\mathrm{sym}}, Z_{\mathrm{sym}}, Z_{\mathrm{sat}}$ remain largely unconstrained. 

This behavior follows from the metamodel expansion of the nuclear energy density, in which the pressure of neutron-rich matter is primarily controlled by the density dependence of the symmetry energy. The slope and curvature parameters, $L_{\mathrm{sym}}$ and $K_{\mathrm{sym}}$, enter at higher orders in the density expansion and are therefore most relevant at densities around $(1–2)n_{\mathrm{sat}}$ probed by tidal deformabilities. Although $Q_{\mathrm{sat}}$ characterizes symmetric nuclear matter, it can still affect the total pressure of $\beta$-equilibrated matter through the symmetric contribution, given the small but nonzero proton fraction in NSs. Hence, this could explain why the data still retain some sensitivity to $Q_{\mathrm{sat}}$. 

\noindent
\textit{Inference with the MM model:} The MM framework yields substantially weaker constraints overall. Even in the combined analysis, only the symmetry energy slope exhibits a mild information gain (JSD increases to about $0.1$), while all other parameters remain weakly constrained as in the single-event analysis. Although the MM set does not impose $\chi$EFT or explicit SS constraints, it is not fully agnostic. The EOS in the MM framework are required to remain causal up to at least $1.8M_{\odot}$, which introduces an extra stiffness condition on $L_{\mathrm{sym}}$, while $K_{\mathrm{sym}}$ gets less negative (cf.~Fig.~\ref{fig:injEOS_basedon_Nucparams}). Overall, we find that the inference of NEPs in the MM framework is largely prior driven, with tidal deformability measurements providing only a mild information gain, leading to the very small JSD values obtained in the MM analysis.

\noindent
\textit{Inference with the Skyrme model:} The Skyrme framework exhibits the strongest overall constraints among the models considered. In this case, the information gain for $L_{\mathrm{sym}}$ increases to $0.27$, representing the largest JSD value in our study. Likewise, $Q_{\mathrm{sym}}$ reaches the highest JSD value of $0.23$ across all EOS sets, while all other NEPs remain weakly constrained in the combined analysis (cf.~Table~\ref{JSDs_all_NUCmodels}). This behavior reflects the fact that the remaining NEPs enter the EOS either at higher orders in the density expansion or through combinations that are weakly correlated with tidal deformabilities, limiting the sensitivity of GW data to these parameters even in the combined analysis.

Because the Skyrme framework also allows consistent calculation of nuclear structure properties, in addition to our inference of NEPs, we examine how the GW-inferred constraints propagate to derived finite-nucleus observables within the Skyrme framework. In particular, we compute neutron skin thickness $\Delta r$ and electric dipole polarizability $\alpha_{D}$, which are known to be sensitive to the density dependence of the symmetry energy (cf.~\cite{vonNeumann-Cosel:2025hsj} for a recent review). The Skyrme Hartree-Fock+BCS+RPA calculations were performed with the code \texttt{skyrme-rpa} \cite{coloSelfconsistentRPACalculations2013}.

\begin{figure}[t]
    \centering
    \includegraphics[width=\columnwidth]{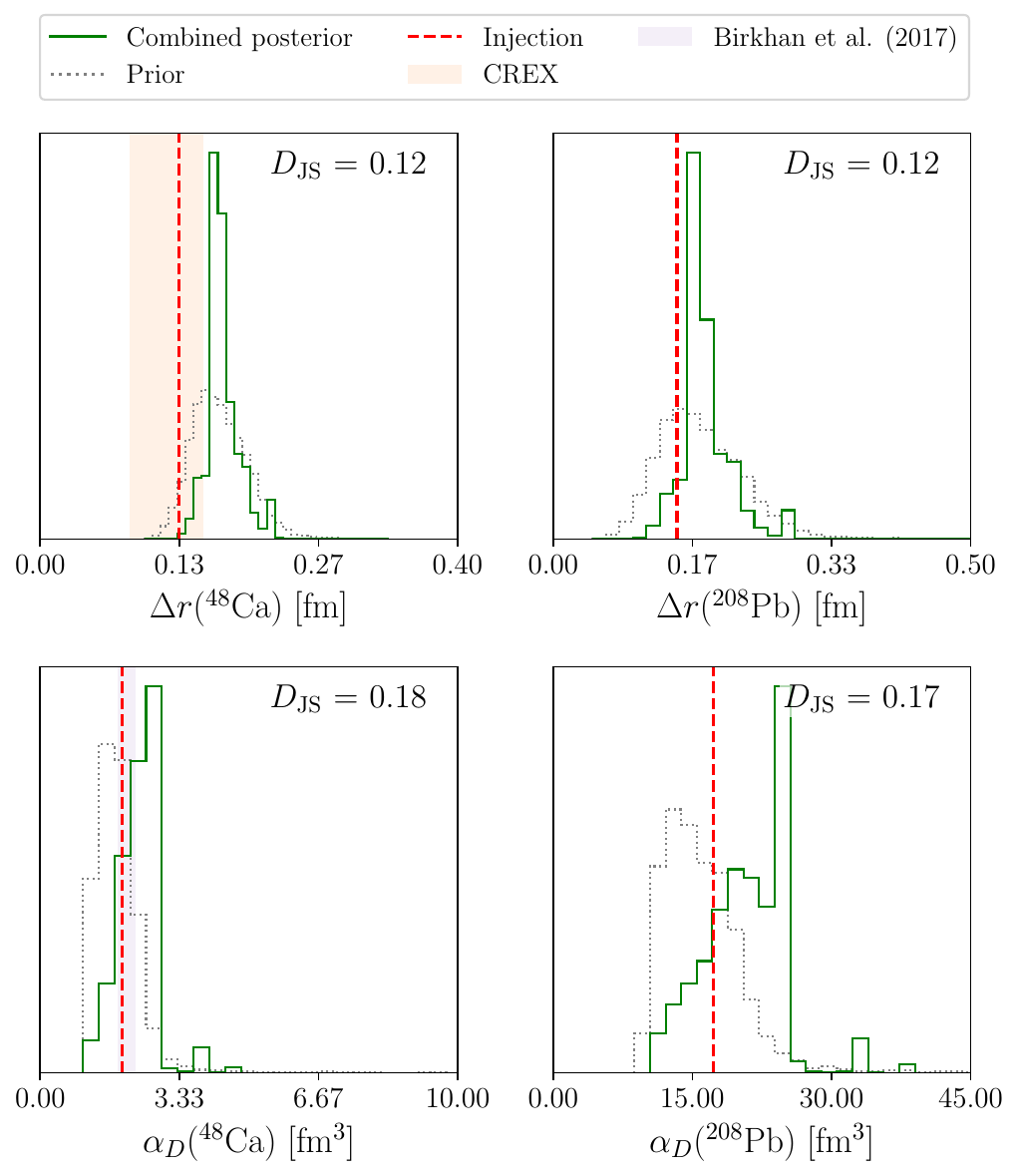}
    \caption{Combined constraints on the neutron skin thicknesses $\Delta r$ and electric dipole polarizabilities $\alpha_{D}$ for ${}^{208}\rm Pb$ and ${{}^{48}\rm Ca}$ obtained from the joint analysis of simulated BNS~events A–D. The combined posterior distributions (green) are compared with the corresponding priors (gray), while the injected values (red dashed lines) correspond to quantities consistently computed within the Skyrme EDF framework from the NEPs of the injected Skyrme EOS (cf.~Fig.~\ref{fig:injEOS_basedon_Nucparams}). The JSD, $D_{\rm JS}$, between the combined posterior and prior is reported in the upper right corner of each panel. Vertical shaded bands with $1\sigma$ ranges denote previous findings: CREX for $\Delta r({}^{48}\mathrm{Ca})$ \cite{CREX:2022kgg} and Birkhan et al.\ (2017) for $\alpha_{D}({}^{48}\mathrm{Ca})$~\cite{Birkhan:2016qkr}.}
    \label{fig:combPOST_dr_alphaD_skyrme}
\end{figure}

The results obtained from the joint analysis of simulated BNS~events A–D are shown in Fig.~\ref{fig:combPOST_dr_alphaD_skyrme} for ${{}^{48}\rm Ca}$ and ${}^{208}\rm Pb$. For reference, the figure also includes experimental results from CREX as well as photoabsorption measurements reported by~\citep{Birkhan:2016qkr}. However, we note that the injected values were not chosen to reproduce these measurements and their consistency arises independent of any experimental calibration. Our results show that there is a moderate information gain for both the neutron skin thicknesses and electric dipole polarizabilities when information from multiple BNS events is combined. The constraining power of the combined posteriors is characterized by JSD values of approximately $0.12$ for the neutron skins of ${{}^{48}\rm Ca}$ and ${}^{208}\rm Pb$, while the electric dipole polarizabilities exhibit a higher degree of information gain at approximately $0.17$--$0.18$. This level of constraint is similar for both ${{}^{48}\rm Ca}$ and ${}^{208}\rm Pb$, suggesting a comparable level of constraint for both nuclei. While the combined posteriors for the neutron skin thicknesses are tighter than the prior, the combined posteriors for the dipole polarizabilities exhibit a shift to larger values for both ${{}^{48}\rm Ca}$ and ${}^{208}\rm Pb$. Moreover, for $\Delta r({{}^{48}\rm Ca})$, the injected value lies outside the main support of the combined posterior. This behavior is consistent with the shift toward larger $L_{\mathrm{sym}}$ values (cf.~Fig.~\ref{fig:Sysbiases_complete}) and reflects the strong correlation between neutron skin thickness and $L_{\mathrm{sym}}$ together with the GW-driven preference for stiffer EOS realizations. An assessment of this bias across all EOS models is presented in the following subsection.

\subsection{Systematic biases from variations of the underlying equation of state}

\noindent
We now examine the impact of EOS model biases on the inference of NEPs by recovering simulated signals with EOS frameworks different from those used for injection. The analysis is performed using the previously selected EOS from MM-SS, MM, and Skyrme sets and the combined-event posteriors are shown in Fig.~\ref{fig:Sysbiases_complete}. 

Across all EOS frameworks, we observe a systematic shift of the inferred symmetry energy slope toward larger values. This may be attributed to several contributing factors. These include the indirect sensitivity of the tidal deformability to $L_{\rm sym}$ through its correlation with NS radii, the influence of prior structure and parameter degeneracies in regimes where the data are only weakly informative, and prior volume effects that favor larger $L_{\rm sym}$ values, particularly in models subject to crust stability constraints. When combining multiple events, such preferences may further accumulate, leading to an apparent shift in the posterior. As a consequence, the injected values of $L_{\mathrm{sym}}$ are consistently overestimated, with posterior peaks typically lying in the range $60$--$80$~MeV, corresponding to an overestimation of approximately $25$--$50$~MeV. 

While the precise magnitude of this shift is difficult to quantify due to the strongly multimodal posterior structure, variations in the posterior peak locations across recovery models indicate a residual EOS-dependent systematic uncertainty at the level of tens of MeV. For example, for the MM-injection scenario, the Skyrme posterior peaks near $90$~MeV, while the main mode for the MM-SS and MM recovery is located roughly between $70$ and $80$~MeV. This discrepancy points to a potential model-dependent systematic uncertainty of approximately $20$~MeV in the inference of the symmetry energy slope. However, we note that this estimate should be interpreted with caution due to the multimodal structure of the inferred posterior distributions. We expect that these result from the discrete and nonlinear mapping between EOS realizations and NEPs employed in our study, in which NEPs are not sampled directly. 

For the symmetry energy curvature $K_{\mathrm{sym}}$, the inferred posteriors are broad and multimodal, with differing prior boundaries across models, precluding a robust assessment of systematic shifts. The isoscalar skewness parameter $Q_{\mathrm{sat}}$ exhibits model-dependent offsets and the injected values are best recovered by MM-SS and MM. In the case of the MM-SS injection, the systematic discrepancy between the maxima of the self-consistent MM-SS posterior and the MM recovery reaches approximately $900$~MeV. A comparable trend is observed for the MM-injection scenario, where the peak separation between the MM-SS and MM distributions is roughly $700$~MeV. These differences are strongly influenced by the disparate prior support across EOS models and should therefore be interpreted with caution.

With regard to the isovector skewness parameter $Q_{\mathrm{sym}}$, the posteriors exhibit strong model-dependent behavior, with the MM-SS framework favoring large positive values and the Skyrme model consistently preferring negative values, accompanied by significant accumulation at the prior boundaries. As a result, $Q_{\mathrm{sym}}$ shows the largest systematic model-dependent divergence in our analysis, with differences between posterior peaks reaching up to $1500$~MeV when comparing the peak of the MM-SS and Skyrme posterior in the first column.

Our analysis in Fig.~\ref{fig:Sysbiases_complete} shows that systematic biases in the inferred NEPs are present, but that a robust quantitative assessment of these biases is hindered by several limiting factors. Although $L_{\mathrm{sym}}$ exhibits comparatively well-converged posterior distributions, it consistently fails to recover the injected value across all injection and recovery scenarios. In contrast, higher-order parameters ($K_{\mathrm{sym}}$, $Q_{\mathrm{sat}}$, $Q_{\mathrm{sym}}$) suffer from pronounced posterior multimodality, disparate prior support, and boundary effects that preclude a reliable and quantitative comparison of systematic offsets. 

\section{Studying constraints on dark matter properties}\label{sec:results_DM}

\subsection{Studying synthetic signals with next-generation detector observations}\label{subsec:simualted_BNS_signals}

\noindent
To assess the potential of next-generation GW detectors to constrain the DM properties, we simulate the BNS events listed in Table~\ref{tab:BNS_events_params}. As before, GW signals of all BNSs are generated by employing the aligned-spin waveform model \texttt{IMRPhenomD\_NRTidalv2} \cite{Dietrich:2019kaq}.\footnote{We performed consistency checks using the more recent waveform model \texttt{IMRPhenomXAS\_NRTidalv3} \cite{Pratten:2020fqn, Abac:2023ujg}, which yielded similar results.} As underlying EOS sets, we have used both the MM-DMA and Skyrme-DMA EOS sets. In each case, we adopt the same baryonic EOS selected in Sec.~\ref{sec:results_BM} and construct injections with the corresponding DM-informed MM or Skyrme EOS set. For all simulated BNS configurations, the DM particle mass is fixed to $m_{\chi} = 483$~MeV. The assumed DM mass fractions vary across the simulated events and are chosen to span a range of small admixtures. BNS event A is injected with a DM fraction of $f_{\chi} = 0.5\%$, BNS event B with $f_{\chi} = 0.04\%$, and BNS events C–D with $f_{\chi} = 0.3\%$. In the following, we present the results of our DM parameter inference for each DMA EOS set and for each GW detector setup.

\begin{figure*}[t]
    \centering
    \includegraphics[width=\textwidth]{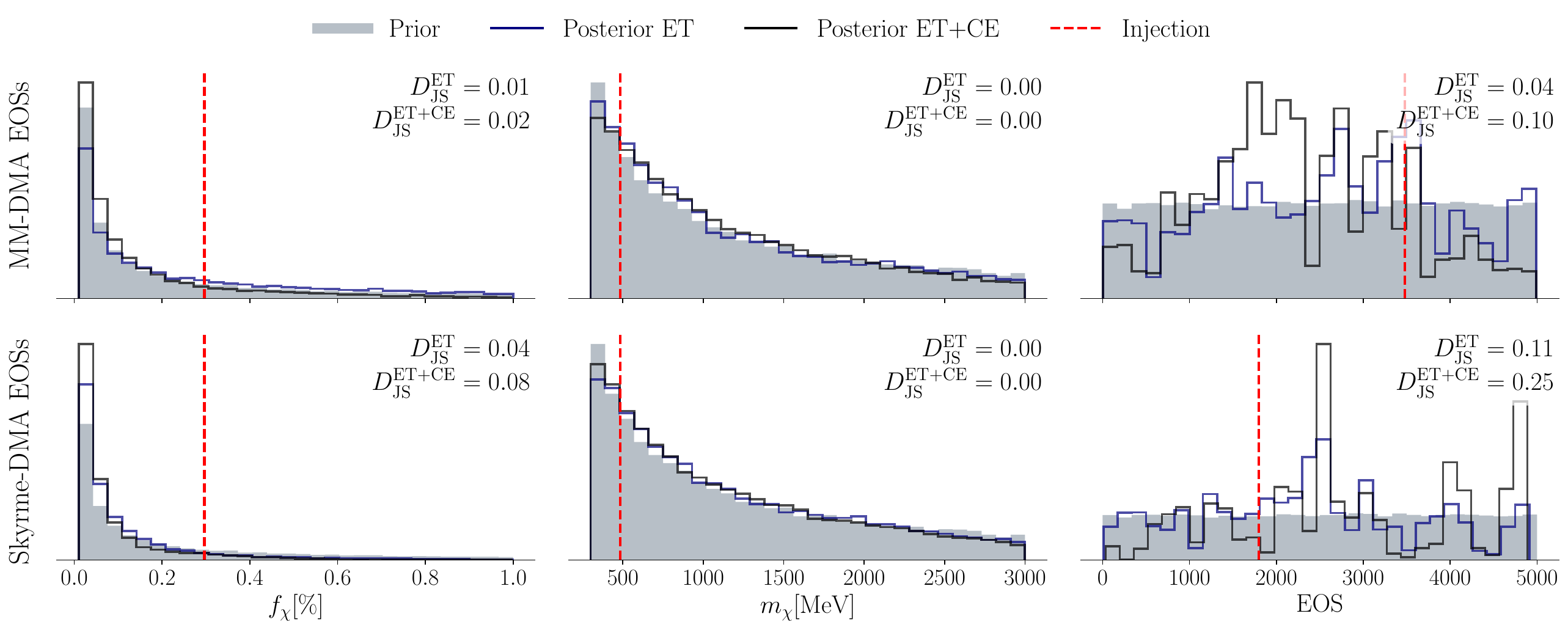}
    \caption[Analysis of the DM constraint for the event~C]{Constraints on DM parameters for BNS event~C. The upper (lower) panels show the parameter-estimation results obtained by sampling with the DMA MM and (Skyrme) EOS set, respectively. In each subpanel, the prior (gray shaded) is compared to posteriors recovered using the ET (blue) and ET+CE (black) detector configurations, with the JSD, $D_{\rm JS}$, quantifying the information gain of the posterior relative to the prior. Red dashed lines indicate the injected values of the DM fraction, particle mass, and EOS index. The DM particle mass is fixed to $m_{\chi} = 483$~MeV and the DM fraction for this simulated BNS event~C is $f_{\chi} = 0.3$\%.}
    \label{fig:DM_constraint_highestSNR_event_ETCE_MMSkyrme}
\end{figure*}

Figure~\ref{fig:DM_constraint_highestSNR_event_ETCE_MMSkyrme} evaluates how well future GW detectors (ET and ET+CE) can narrow down the DM fraction, DM particle mass, and the EOS of a given underlying baryonic EOS. Figure~\ref{fig:DM_constraint_highestSNR_event_ETCE_MMSkyrme} shows the constraints obtained for the simulated BNS event~C with the highest SNR. For both the MM and Skyrme DMA EOS sets, the recovered posterior distributions for the DM fraction and particle mass resemble the initial prior distributions, and the JSD values for these parameters remain negligible. Due to the degeneracies between the unknown baryonic EOS and effects introduced through the presence of DM, the obtained posteriors on the DM particle mass remain, as expected, uninformative regardless of the detector configuration.

These results contrast with the recovery of the EOS index, which shows a substantially higher information gain across all scenarios and is consistent with the enhanced EOS sensitivity expected for third-generation GW detectors. Although the posterior distributions for the EOS index show multimodal structures, the JSD values are notably higher, reaching $0.25$ for the Skyrme-DM EOS when using the ET+CE network.

In addition to assessing whether DM properties can be constrained, we have examined the robustness of the parameter recovery under both the BM and DMA assumptions. In particular, we analyze whether injecting signals generated with DMA compositions leads to systematic biases when the data are recovered under either the BM or DMA hypothesis within the MM and Skyrme frameworks. We do not observe significant systematic offsets in the recovered posteriors for either scenario. For both detector configurations, ET and ET+CE, the inferred source parameters, including chirp mass and tidal deformability, remain consistent within their respective statistical uncertainties. This behavior is consistent across both EOS frameworks considered.

Overall, these results indicate that the discriminating power of tidal deformability measurements is limited by EOS degeneracies. This finding is consistent with our observation that no significant information gain on the considered DM parameters is obtained, irrespective of the assumed EOS or GW~detector configuration considered. 

\subsection{A unified analysis: Impact of dark matter on nuclear inference}\label{sec:DMbias_on_NEPs}

\noindent
To assess the impact of a potential DM component on the inference of NEPs, we compare the NEP posteriors obtained when recovering the signals with DMA EOS to those obtained using the corresponding BM EOS. The difference between the two posterior distributions is quantified using the JSD. In this context, the JSD provides a quantitative measure of how strongly the inferred NEPs differ between the DMA and baryonic scenarios, thereby characterizing the potential bias induced by neglecting a DM component in the analysis. The results are reported in Table~\ref{tab:JSD_results_NEPs_DMpresence} for the models (MM and Skyrme) and the detector configurations (ET and ET+CE) and quantify the systematic bias introduced by DM admixtures. 

\begin{table}[h!]
\centering
\renewcommand{\arraystretch}{1.2} 
\begin{tabular}{|l| >{\centering\arraybackslash}p{1.5cm}| >{\centering\arraybackslash}p{1.5cm}| >{\centering\arraybackslash}p{1.5cm}| >{\centering\arraybackslash}p{1.5cm}|}
\hline
\multirow{3}{*}{\textbf{NEP}} & \multicolumn{4}{c|}{\textbf{JSD $D_{\rm JS}$ (in bits)}} \\ \cline{2-5} 
 & \multicolumn{2}{c|}{\textbf{MM}} & \multicolumn{2}{c|}{\textbf{Skyrme}} \\ \cline{2-5} 
 & \textbf{ET} & \textbf{ET+CE} & \textbf{ET} & \textbf{ET+CE} \\ \hline
$E_{\rm{sym}}$ & 0.01 & 0.06 & 0.01 & 0.02 \\ 
$L_{\rm{sym}}$ & 0.01 & 0.02 & 0.01 & 0.02 \\ 
$K_{\rm{sat}}$ & 0.01 & 0.01 & 0.01 & 0.02 \\ 
$K_{\rm{sym}}$ & 0.02 & 0.03 & 0.01 & 0.04 \\ 
$Q_{\rm{sat}}$ & 0.01 & 0.02 & 0.01 & 0.03 \\ 
$Q_{\rm{sym}}$ & 0.01 & 0.02 & 0.01 & 0.04 \\ 
$Z_{\rm{sat}}$ & 0.02 & 0.01 & 0.01 & 0.11 \\ 
$Z_{\rm{sym}}$ & 0.02 & 0.01 & 0.03 & 0.12 \\ \hline
\end{tabular}
\caption{Impact of DM on the inference of NEPs. JSD values (in bits) quantify the discrepancy between combined posteriors of NEPs obtained using pure BM versus DMA BNS merger simulations for both the MM and Skyrme models under ET and ET+CE detector configurations.}
\label{tab:JSD_results_NEPs_DMpresence}
\end{table}

The inferred JSD values listed in Table~\ref{tab:JSD_results_NEPs_DMpresence} are small, suggesting that, if DM is present, its impact on the inference of NEPs within the considered models and GW detector configurations is expected to be small. The transition from a single- (ET) to a dual-network configuration (ET+CE) consistently leads to higher JSD values, particularly for the Skyrme EOS. While both the MM and Skyrme models show low levels of bias at ET sensitivity, the Skyrme model appears more susceptible to inference bias in high-order parameters when high-precision (ET+CE) data are available. However, these biases can be attributed to more pronounced peaks in their multimodal posterior structure. A similar behavior is observed for the MM, for which $E_{\rm sym}$ achieves the highest JSD value of $0.06$ under the ET+CE detector configuration, arising from differences in the shape of its multimodal posterior rather than from a coherent shift in the inferred parameter.

Although our JSD results exhibit some dependence on the assumed EOS and GW detector configuration, the differences between combined NEP posteriors obtained from pure BM versus DMA BNS mergers remain negligible. In addition, the assessment of systematic shifts is complicated by the inherently multimodal posterior structure of several NEPs, which limits the interpretability of apparent divergences. Overall, these results indicate that, within the scope of the present analysis, potential biases induced by DM admixtures in BNS mergers on the inference of NEPs are negligible. 

\section{Discussion}\label{Sec:Dicussion}

\subsection{Scope and robustness of the analysis}

\noindent
Before discussing the physical implications of our results, we outline the assumptions adopted in this work and their potential influence on the final results.

First, our Bayesian inference framework relies on precomputed EOS sets for both the inference of NEPs and DM properties. For NEP inference, direct sampling of NEPs would allow a more continuous exploration of parameter space and could reduce discretization-induced multimodality, potentially yielding smoother posteriors and modestly increased information gain (cf.~\cite{Reed:2024urq, Wouters:2025zju}). Nevertheless, as shown before, e.g., in~\cite{Mondal:2021vzt,Iacovelli:2023nbv}, strong intrinsic degeneracies among NEPs limit the extent to which direct sampling alone can improve constraints from GW observations. Similarly, for DM inference, we rely on precomputed DMA EOS grids spanning a finite range of DM particle masses and fractions. Direct sampling over DM parameters and the EOS would mitigate discretization effects, but is not expected to alleviate degeneracies between DM-induced effects and EOS uncertainties at the small DM fractions considered here.

Second, we simulated DMA BNS merger signals using existing GW models, which do not incorporate DM effects directly. While waveform models including DM-induced modifications could reduce modeling uncertainties and improve physical consistency, the small DM fractions considered here imply that DM effects remain largely degenerate with EOS variations, such that the overall impact on DM inference is expected to be limited.

Third, our analysis was restricted to a specific class of DM candidates, namely non-self-interacting fermionic DM coupled to baryonic matter only through gravity. Other DM scenarios, such as bosonic DM, may induce different effects on NS structure and hence on the GW signal. 

\subsection{Nuclear empirical parameter inference: Constraints and systematic effects}

\noindent
Our study confirmed that only weak constraints on the NEPs can be obtained from single BNS observations, while combining multiple events can, in principle, enhance the constraining power. In the combined analysis, we find moderate information gain for $L_{\rm sym}$ and, depending on the EOS, for $K_{\rm sym}$, $Q_{\rm sym}$, and $Q_{\rm sat}$. Overall, the constraints are strongly model dependent, with the Skyrme framework yielding the tightest constraints. This shows that EOS modeling choices play a critical role in the interpretation of NEP constraints with next-generation GW detectors. Importantly, our results show that systematic shifts are already present when injection and recovery are performed within the same EOS framework, complicating a reliable quantification of systematic effects. 

Our analysis also showed that inferred NEPs can exhibit substantial systematic offsets when the recovery EOS differs from the injected one. Most notably, $L_{\rm sym}$ is systematically biased toward larger values across all injection and recovery scenarios. This behavior indicates that tidal information, combined with EOS-specific correlations and prior restrictions, can favor effective symmetry energy slopes that differ from the injected value. For higher-order NEPs, the assessment of systematic shifts is hindered by limited information gain, posterior multimodality, and intrinsic parameter degeneracies. 

\subsection{Dark matter in neutron stars: Degeneracies and limitations}

\noindent
In Sec.~\ref{sec:results_DM}, we examined whether the presence of a DM component in neutron stars can be constrained from GW observations. Across both EOS frameworks (MM and Skyrme) and detector configurations (ET and ET+CE), the posterior distributions of the DM particle mass and fraction presented in Fig.~\ref{fig:DM_constraint_highestSNR_event_ETCE_MMSkyrme} remain largely consistent with their priors, indicating no significant information gain within the scope of our analysis. These findings are consistent with the expectation that, for the small DM fractions considered in this study, DM-induced modifications to tidal deformabilities can be strongly degenerate with EOS variations, limiting the ability of GW~observations with ET alone to isolate DM effects. Although the DM-induced modifications in the tidal deformability could be measurable by ET for the highest SNR events (events A and C) considered in this study, these shifts are largely ``absorbed'' through degeneracies between the influence of DM and the unknown EOS. As a consequence, even in favorable cases, the identification of a DM component from tidal information alone remains challenging without additional constraints or complementary observations. Our results show that this conclusion remains unchanged even with the inclusion of CE. However, for the class of DM models considered here, i.e., non-self-interacting fermionic DM in core configurations with mass fractions below 1\%, the impact of a neglected DM component on the inferred NEPs is found to be negligible compared to the dominant EOS-driven uncertainties encountered in standard non-DM NS analyses.

\section{Summary}\label{sec:conclusion}

\noindent
In this work, we assessed the ability of next-generation GW detectors to constrain nuclear physics and dark matter properties using BNS mergers.

We find that, despite the sensitivity of next-generation GW detectors, constraints on NEPs from BNS mergers remain limited and highly model dependent, underscoring the critical role of EOS modeling choices and the need to carefully account for EOS-induced systematic uncertainties. Our results indicate that, for non-self-interacting fermionic DM in core configurations with mass fractions below $1\%$, GW observations alone are unlikely to yield robust constraints on DM properties or to unambiguously identify DMA NSs, as DM-induced tidal signatures remain strongly entangled with EOS uncertainties, even in the high-SNR regime and with next-generation detector networks. 

Overall, these findings highlight both the promise and the limitations of GW~observations as probes of dense matter microphysics. Realistic progress will likely depend on combining GW data with complementary astrophysical observations, improving microphysical modeling to better control intrinsic NEP degeneracies, refining waveform models, and developing inference frameworks that can account for nuclear and DM physics effects in a consistent way.

\acknowledgements

\noindent
N.~K. acknowledges support by the ANR GalaxyFIT project, Grant No. ANR-24-CE31-5542 of the French Agence Nationale de la Recherche. V.~S. gratefully acknowledges support from the UKRI-funded ``The next-generation gravitational-wave observatory network'' project (Grant No. ST/Y004248/1).
This work was cofunded by the European Union (ERC, SMArt, 101076369). Views and opinions expressed are those of the authors only and do not necessarily reflect those of the European Union or the European Research Council. Neither the European Union nor the granting authority can be held responsible for them.

\section*{Data Availability}

\noindent
There are no publicly available research data or software supporting this manuscript. Requests for further information or data should be sent to the authors.

\appendix

\section{Construction of NS EOS}\label{App:EOS_construction}

\noindent
Here, we provide a detailed description of how the EOS were constructed, given the three models considered in this study. 

\textbf{(i) Metamodel approach with speed-of-sound extension:} We employ the EOS set presented in previous studies~\cite{Komoltsev:2023zor, Koehn:2024set} based on the MM approach \cite{Margueron:2017eqc, Margueron:2017lup}. Here, we briefly recap the main construction steps and further details can be found in~\cite{Koehn:2024set}. All EOS in this set use the crust model from~\cite{Douchin:2001sv}, which is applied up to the crust-core transition density at $0.076$~fm$^{-3}$. The EOS of the outer NS core assumes nucleonic matter in $\beta$-equilibrium up to a transition density $n_{\rm break}$, randomly drawn from $(1–2)~n_{\rm sat}$. Above this density, we effectively account for the possibility of a deconfinement phase transition to quark matter~\cite{Somasundaram:2021clp}. Below $n_{\rm break}$, the MM of \cite{Margueron:2017eqc, Margueron:2017lup} is used, a density functional approach incorporating nuclear physics knowledge through NEPs defined via a Taylor expansion of the energy per particle in symmetric matter $e_{\rm sat}$ and the symmetry energy $e_{\rm sym}$ around saturation density $n_{\rm sat}$. 

To account for nuclear physics uncertainties and to generate a broad EOS prior, NEPs are varied uniformly within the ranges in Table~\ref{tab:NEP_range_MM}. This defines the EOS in the range $0.12~\mathrm{fm}^{-3}~<~n~<~n_{\text{break}}$, with the lower limit chosen arbitrarily. Above $n_{\rm break}$, an agnostic approach based on a modified SS parametrization is incorporated (originally introduced in~\cite{Tews:2018iwm}, see~\cite{Koehn:2024set} for details). This framework allows for arbitrarily soft EOS that mimic the first-order phase transitions. We refer to this EOS family as ``MM-SS.''

Following this methodology, a set of 100,000 EOS was generated and utilized in previous studies, including~\cite{Komoltsev:2023zor, Koehn:2024set}. Unless stated otherwise, most analyses are conducted on a subset presented in~\cite{Koehn:2024set} to reduce computational costs, which excludes EOS that are inconsistent with the most reliable constraints (cf.~references in~\cite{Koehn:2024set}), such as heavy pulsar mass measurements via radio timing and theoretical predictions from $\chi$EFT and perturbative quantum chromodynamics~\cite{Kurkela:2009gj}.

\begin{table}
\centering
\tabcolsep=0.5cm
\def\arraystretch{1.5}
\begin{tabular}{>{\centering\arraybackslash}p{1.8cm} >{\centering\arraybackslash}p{1.5cm} >{\centering\arraybackslash}p{1.8cm}}
 \hline
 \hline
 \textbf{Parameter} & \multicolumn{2}{c}{\textbf{Prior}} \\
  & MM-SS & MM \\
 \hline
 $n_{\rm break}$ [$n_{\rm sat}$] & $\mathcal{U}(1,2)$  & - \\
 $E_{\rm sat}$ [MeV] &                  -           &  $\mathcal{U}(-17,-14.5)$\\
 $n_{\rm sat}$ [fm$^{-3}$] &            -                & $\mathcal{U}(0.14, 0.17)$\\
 $K_{\rm sat}$ [MeV] & $\mathcal{U}(150,300)$     & $\mathcal{U}(130,300)$\\
 $Q_{\rm sat}$ [MeV] & $\mathcal{U}(-500,1100)$   & $\mathcal{U}(-900,900)$\\
 $Z_{\rm sat}$ [MeV] & $\mathcal{U}(-2500,1500)$  & $\mathcal{U}(-1000,1000)$ \\
 $E_{\rm sym}$ [MeV] & $\mathcal{U}(28,45)$       & $\mathcal{U}(28,60)$ \\
 $L_{\rm sym}$ [MeV] & $\mathcal{U}(10,200)$      & $\mathcal{U}(25,160)$ \\
 $K_{\rm sym}$ [MeV] & $\mathcal{U}(-300,100)$    & $\mathcal{U}(-300,200)$ \\
 $Q_{\rm sym}$ [MeV] & $\mathcal{U}(-800,800)$    & $\mathcal{U}(-800,800)$ \\
 $Z_{\rm sym}$ [MeV] & $\mathcal{U}(-2500, 1500)$ & $\mathcal{U}(-1000,1000)$ \\
\hline
\hline
\end{tabular}
\caption{The prior distributions for NEPs used for the construction of both EOS sets: MM-SS and MM. Uniform priors are denoted by $\mathcal{U}$. For MM-SS, the parameters $E_{\rm sat}$ and $n_{\rm sat}$ are fixed at $-16$~MeV and $0.16$~fm$^{-3}$, respectively.}
\label{tab:NEP_range_MM}
\end{table}

\textbf{(ii) Metamodel approach with unified crust-core EOS (MM):} 
We generate a second set of roughly 35,000 EOS using the MM~\cite{Margueron:2017eqc} approach mentioned in (i). There are two main differences from the previous set: (a) Here, we construct a unified EOS for the NS crust and core. The crust is modeled with the compressible liquid drop model as presented in~\cite{Grams:2022a}, where the underlying NEPs of the crust are the same as in the NS core. (b) We keep only the MM at the core. This construction assumes the nucleonic hypothesis and uses the MM at all densities. Therefore, the EOS corresponding to the NEPs is used to model both the inner and outer core. We refer to this EOS family as ``MM.''

\textbf{(iii) Skyrme model:} We generate a third set of EOS using nonrelativistic density functional theory. We utilize an extended Skyrme EDF~\cite{Davesne:2015lca,zhangExtendedSkyrmeInteractions2016} with a set of density-dependent terms given by
\begin{equation}
    \mathcal{H}_{\rho} = \sum_{i=3}^6 \frac{1}{4} t_{i} \rho^{2+\alpha_i} [(2+x_{i}) - (2x_{i}+1)(y_{p}^{2}+y_{n}^{2})].
\end{equation}

The exponents of these density-dependent terms $\alpha_i$ take the values ${\frac{1}{3}}, {\frac{2}{3}}, 1, {\frac{4}{3}}$ reflecting an expansion in powers of Fermi momentum \cite{limStructureNeutronStar2017}. The parameters of the symmetry energy $E_{\rm sym}$, $L_{\rm sym}$, $K_{\rm sym}$ and $Q_{\rm sym}$ and symmetric matter EOS $n_{\rm sat}$, $E_{\rm sat}$, $K_{\rm sat}$, $Q_{\rm sat}$ can be written in terms of the remaining eight coefficients of the density-dependent term $x_{3-6}$, $t_{3-6}$~\cite{Xu:2022tod}. Using these EDFs, we construct a unified crust-core EOS using the compressible liquid drop model for the inner crust and the Baym-Pethick-Sutherland EOS~\cite{Baym:1971pw} for the outer crust, whose composition is relatively well determined by currently available nuclear masses. Details and applications of these EOS can be found in~\cite{Newton:2011dw,Newton:2021rni,Balliet:2020nsh}.

We generate 17,000 EOS starting from the same uniform priors on the nuclear matter parameters as the metamodel \emph{except} for the ranges for $E_{\rm sym}$ and $L_{\rm sym}$ which are $\mathcal{U}(25, 45)$ and $\mathcal{U}(10, 150)$, respectively, and further filter for those EOS which reach a maximum mass of $1.99~M_{\odot}$ to cover the high component masses of BNS event D (cf.~Table~\ref{tab:BNS_events_params}). The final Skyrme EOS set shown in Fig.~\ref{fig:MRL_EOSsets} includes roughly $7000$~EOS.

In addition to the slightly smaller parameter space explored for the $E_{\rm sym}$ and $L_{\rm sym}$ parameters, this model contains slightly less flexibility than the MM to explore the highest density regime of the EOS because the fourth-order nuclear matter parameters $Z_{\rm sat}$ and $Z_{\rm sym}$ are not independently variable and are correlated with the other NEPs.

An advantage of using an energy-density functional is that we can use the model to consistently calculate the neutron skins of $^{208}$Pb and $^{48}$Ca within the Skyrme-Hartree-Fock theory, enabling us to consistently make a link between the posteriors we obtain and the experimental nuclear measurements of neutron skins~\cite{PREXII,CREX:2022kgg} and electric dipole polarizabilities \cite{vonNeumann-Cosel:2025hsj,Tamii:2013cna}. We use the code \texttt{skyrme-rpa} \cite{coloSelfconsistentRPACalculations2013} to perform such calculations.

\bibliography{main.bib}

\end{document}